\magnification=1200
\baselineskip=20pt
\raggedbottom
\interlinepenalty=200
\hsize=6.5truein
\vsize=9.0truein
\voffset=0.0truein
\hoffset=0.0truein
\parindent=0pt
\parskip=0pt

\def\smallskip{\vskip 2pt}
\def\medskip{\vskip 12pt}
\def\bigskip{\vskip 48pt}

\def\page{\vfill\eject}
\centerline{ }
\bigskip
\centerline{\bf RAPID FORMATION OF ICE GIANT PLANETS}
\bigskip
\centerline{Alan P.~Boss, George W.~Wetherill, and Nader Haghighipour}
\medskip
\centerline{\it Department of Terrestrial Magnetism}
\centerline{\it Carnegie Institution of Washington}
\centerline{\it 5241 Broad Branch Road, N.W., Washington, D.C. 20015-1305}
\centerline{(202)478-8858, FAX (202)478-8821, boss@dtm.ciw.edu}
\vskip 100pt
\centerline{Number of manuscript pages: 20}
\centerline{Number of figures: 0 }
\centerline{Number of tables: 1 }
\medskip
\medskip
\centerline{[Submitted as a {\it Note} to {\it Icarus}]}

\page
\centerline{ }
\vskip 100 pt
\centerline{Suggested running head: Ice Giant Planet Formation}
\vskip 200 pt
\centerline{Send correspondence to: }
\medskip
\centerline{Dr. Alan P. Boss }
\centerline{Department of Terrestrial Magnetism }
\centerline{Carnegie Institution of Washington }
\centerline{5241 Broad Branch Road, N.W. }
\centerline{Washington, D.C. 20015-1305 }

\page
\parindent=20pt
\parskip=0pt
\centerline{{\bf ABSTRACT}}
\medskip

 The existence of Uranus and Neptune presents severe difficulties 
for the core accretion model for the formation of ice giant 
planets. We suggest an alternative mechanism, namely disk instability 
leading to the formation of gas giant protoplanets, coagulation and
settling of dust grains to form ice/rock cores at their centers, and
photoevaporation of their gaseous envelopes by a nearby OB star, as 
a possible means of forming ice giant planets. 

\medskip
\medskip
\centerline{Key words: Neptune; planetary formation; Uranus}
\medskip

\page
\centerline{\bf INTRODUCTION}
\medskip

 The robustness of the current theory of the origin of the Solar System
deteriorates sharply with increasing distance from the Sun. The
innermost, terrestrial planets are widely believed to have formed
by the collisional accumulation of rocky planetesimals, culminating
after $\sim$ 50 Myr in giant impacts between planetary-sized bodies 
(Canup and Righter 2000). The gas giant planets, Jupiter and Saturn, 
were formed by core accretion in the conventional view, where a 
$\sim 10$ Earth-mass ($\sim 10 M_\oplus$) solid core forms 
first by collisional accumulation, and then accretes a gaseous
envelope from the solar nebula, a process requiring an estimated 8 Myr
in the standard model (Pollack {\it et al.} 1996). However, most young 
stars lose their gas disks in a few Myr or less in regions of low-mass 
star formation (Brice\~no {\it et al.} 2001) and on even shorter time 
scales in OB associations (Bally {\it et al.} 1998), meaning that core 
accretion can only produce gas giant planets in the rare long-lived disk. 
Alternatively, gas giant protoplanets might form rapidly (in $\sim 10^3$ yrs)
by a disk instability, where the gas in a marginally gravitationally
unstable disk forms clumps which then contract to planetary densities 
(Boss 1998, 2000, 2001a). Disk instability operates fast enough to form
gas giant protoplanets in even the shortest-lived protoplanetary disks.

 The problems facing the formation {\it in situ} of the ice giant planets, 
Uranus and Neptune, by the core accretion mechanism are even 
more severe than those facing the formation of Jupiter and Saturn.
If the ice giants formed {\it in situ}, the lower surface
density of solids and longer orbital periods require that 
collisional accumulation proceed much slower than in the gas giant planet
region (inside $\sim$ 10 AU). Even more importantly, because the escape 
velocity from the Solar System at 20 AU to 30 AU (the approximate orbital
distances of Uranus and Neptune, respectively) is $\sim$ 8 km s$^{-1}$,
comparable to orbital velocities and to the relative velocities between 
growing embryos, the effect of mutual encounters is to excite orbital 
eccentricities so much that the embryos cross the orbits of Saturn or 
Jupiter, and thus are either ejected on hyperbolic orbits,
lost by impact with the gas giant planets, or kicked into cometary
orbits (Lissauer {\it et al.} 1995). Ice giant planets cannot form in the 
standard model (Levison and Stewart 2001). Possible means for salvaging 
the collisional accretion mechanism include (Levison and Stewart 2001)
invoking some sort of drag force to damp orbital eccentricities, 
or runaway accretion of a single embryo to Uranus-size (Bryden {\it et al.} 
2000) rather than oligarchic growth of multiple embryos 
(Kokubo and Ida 1998). Perhaps the best suggestion for forming the ice
giant planets by collisional accretion is the proposal that the ice
giants were formed between Jupiter and Saturn and then were gravitationally 
scattered outward (Thommes {\it et al.} 1999).
However, the success of this solution depends 
largely on the ability of core accretion to form gas giant 
planets prior to the removal of the disk gas, which seems to be
limited, considering what we know about gas disk lifetimes.

 Here we propose a completely different mechanism for the formation
of the ice giant planets, motivated by recent theoretical work on gas 
giant protoplanet formation by the disk instability mechanism
(Boss 1998, 2000, 2001a) and by observations of the erosion of 
Orion Nebula protoplanetary disks by extreme ultraviolet (EUV) irradiation 
(Bally {\it et al.} 1998). In a nutshell, we propose forming the
ice giant planets rapidly through the sedimentation of solids in massive
gas giant protoplanets formed by disk instability, followed by removal 
of most of their gaseous envelopes by intense EUV irradiation from a 
nearby massive star. We believe that this mechanism may represent a
superior means for forming ice (and gas) giant planets in the majority
of protoplanetary disks. Terrestrial planet formation is still
likely to occur even when gas giant planets are formed quickly by a
disk instability (Kortenkamp and Wetherill 2000), and may even 
be facilitated (Kortenkamp {\it et al.} 2001).

\medskip
\centerline{\bf DISK INSTABILITY}
\medskip

 The disk instability mechanism is capable (Boss 2000) of forming
multiple-Jupiter-mass clumps in a marginally gravitationally unstable
disk with a surface density at 5 AU comparable to that believed
necessary to form Jupiter by core accretion. Extremely high spatial
resolution is necessary in order for self-gravitating clumps to
form and to persist; three dimensional (3D) calculations with {\it at least} 
$10^6$ grid points seem to be necessary (Boss 2000), which should be 
compared to the 25,000 particles used by Laughlin and Bodenheimer (1994) 
in their 3D smoothed particle hydrodynamics (SPH) disk simulation.
Disk instability appears to be capable of forming self-gravitating clumps
even when a rigorous thermodynamical description (energy equation,
radiative transfer in the diffusion approximation, detailed equations
of state, and dust grain opacities) of the disk is included (Boss 2001a),
because the disk cooling time is comparable to the relevant dynamical
time scale, namely the orbital period. Models where artificial viscosity 
is used to heat the gas, on the other hand, tend to damp the growth
of nonaxisymmetry (Nelson {\it et al.} 2000, Pickett {\it et al.} 2000a,b).
Artificial viscosity was not employed in the 3D models of Boss (1998, 2000,
2001a). 

 In spite of these promising results, it must be admitted that the
present disk instability models cannot yet be considered definitive,
because of inherent limitations, such as the spatial resolution employed even
in the highest spatial resolution models to date (Boss 2000, 2001a), and doubts
regarding the long-term survival of the clumps. However, models currently
in progress by one of us (APB) suggest that with specialized techniques
to improve the spatial resolution of, e.g., the Poisson solver for
the gravitational potential, the disk instability mechanism becomes 
increasingly robust. 

 While the Boss (2000, 2001a) models were limited to clump formation inside 
20 AU, disk instability models with greater radial extent imply that
clumps could still form at somewhat greater distances in a suitable
disk (Nelson {\it et al.} 1998; see Boss 2001a for a discussion of the 
validity of the ``locally isothermal'' approximation used by Nelson 
{\it et al.} 1998 and Boss 2000, compared to the approximations used in 
modeling heating and cooling processes by Nelson {\it et al.} 2000).
The type of protoplanetary disk that would be necessary 
to produce clumps at 20 AU to 30 AU, i.e., a disk with a gravitational 
stability parameter $Q \le 1.5$ in that region, would have a total mass 
inside 30 AU of $\sim$ 0.13 solar masses ($M_\odot$), with $\sim 0.04 M_\odot$ 
residing between 20 AU and 30 AU, based on extending previous disk 
instability models (Boss 2000, 2001a) to a radius of 30 AU.
The assumed result of such a disk model, four gas giant protoplanets
spaced from $\sim$ 5 AU to $\sim$ 30 AU, is plausible based on the 
previous modeling, but has not as yet been calculated convincingly.

\medskip
\centerline{\bf CORE FORMATION}
\medskip

 Table 1 lists the current estimates (Guillot 1999, Hubbard 1992)
for the total, core, and envelope masses for the gas and ice giant planets, 
where the gaseous envelope mass is assumed to be the difference between 
the total mass and the inferred core mass. Note that a range of core masses 
is possible for Jupiter and Saturn, and the midpoint of this range is used 
in Table 1; the core masses of Uranus and Neptune are similarly uncertain, 
so nominal values are employed. Also shown in Table 1 is the 
mass of the gas giant protoplanet that would be needed to account for the
inferred core masses, assuming a gas to dust mass ratio of 50:1 and
that all the dust grains coagulated together and sedimented to the center
of the protoplanet. The final row in Table 1 then lists the amount
of hydrogen and helium gas that would need to be removed to leave
behind a planet with the observed total mass. We do not consider here
the question of the heavy metal enrichment of the envelopes of the
gas giant planets, as for either core accretion or disk instability,
this envelope enrichment is likely to be associated at least in part with
the ingestion of planetesimals after the planet has formed.

\smallskip
\centerline{\it Insert Table 1}
\medskip

 Clump formation can occur very rapidly in a disk instability, on the dynamical
time scale of the orbital period, or within about a few hundred years 
at an orbital distance of 25 AU. Once a well-defined clump forms,
the dust grains within the clump will begin to sediment toward
the center of the clump. Dust grain growth and sedimentation to
the disk midplane may have already begun at the time when the clumps 
formed, giving a head start to this process, but to be conservative, 
we assume that no growth has yet occurred when the clumps form. 
Using the standard approach for calculating the growth of dust grains 
by differential settling in the solar nebula (Weidenschilling 1988),
specialized to the case of a spherical protoplanet, we obtain a 
characteristic growth time (Boss 1998) of

$$ \tau_a = { c_s \over \pi f \rho_p G R}, $$

\noindent
where $c_s = c_s(T)$ is the sound speed, $f$ is the dust to gas mass ratio,
$\rho_p$ is the density of the protoplanet, $G$ is the gravitational 
constant, and $R$ is the radius of the protoplanet. For nominal values at 25 
AU of the gas temperature $T = 50 K$, $f = 0.02$, $\rho_p = 10^{-9}$ g 
cm$^{-3}$, and $R = 2$ AU, we obtain $\tau_a \approx$ 10 yrs. The time to 
grow from size $a_0$ to $a$ is $\sim 3 \tau_a ln(a/a_0)$, so growing
from $a_0 =$ 0.1 $\mu$m to 1 cm-size requires $\sim 10^3$ yrs.
Cm-sized particles can then sediment to the center of the protoplanet
in a time

$$ \tau_c = { 3 c_s \over 4 \pi a G \rho_a },$$

\noindent
where $\rho_a \approx 3$ g cm$^{-3}$ is the density of a dust grain.
For the nominal values, the sedimentation time of cm-sized grains
is $\tau_c \sim 2 \times 10^3$ yrs. Hence the solids inside a gas protoplanet
at $\sim 25$ AU should coagulate and form a central core on a time scale of 
$\sim 3 \times 10^3$ yrs.

 This estimate of the time scale for the formation of a solid core
assumes that the protoplanet is in radiative equilibrium rather
than in convective equilibrium. Wuchterl {\it et al.} (2000)
calculated that a gas giant protoplanet, starting from conditions which 
approximately characterize its formation during a disk instability, is likely 
to become nearly fully convective within about 100 yrs. Turbulent motions 
driven by convection may thus need to be taken into account in assessing 
whether or not dust grains can settle to the center of a protoplanet.
While this situation is clearly deserving of further study, we can
suggest two possible outcomes here. First, considerable dust grain
coagulation in the solar nebula may have already occurred by the time 
at which the disk instability occurs, so the assumption of starting
from 0.1 $\mu$m-size may be too severe. All that is required is
for the solids to be small enough to remain coupled to the gas sufficiently
that they are incorporated into the gaseous clumps during the disk 
instability phase. If, for example, 10-cm-sized solids had formed
prior to the instability, they would settle to the center of the 
protoplanet in a few 100 yrs, perhaps prior to the onset of
widespread convection. Alternatively, detailed calculations of
the coagulation and vertical settling of solids in the solar nebula
show that even in the presence of moderate turbulence (e.g., a turbulent
velocity of 0.01 km s$^{-1}$), settling to the midplane at 1 AU can
occur in about 2000 yrs (see Fig. 9 of Weidenschilling and Cuzzi 1993),
i.e., in about the same time as for a non-turbulent nebula (see their Fig. 
6). These calculations suggest that even a fully convective protoplanet
need not present an insurmountable barrier to core formation. Afterall,
for every updraft there is a downdraft, in isotropic turbulence. It remains
to be seen if the core formation process can produce a core which is
nearly hydrogen-free, as would be required to produce Uranus and
Neptune, and if a hydrogen- and helium-rich envelope will be stable during
and after this process. Detailed radiative hydrodynamical models
will be necessary to address these points (e.g., Wuchterl 1993).

 Water ice is stable for temperatures less than 180 K (Pollack {\it
et al.} 1994) at a density of $10^{-8}$ g cm$^{-3}$. By comparison,
a newly-formed gas giant protoplanet may have a maximum temperature of
about 130 K at a density of $10^{-8}$ g cm$^{-3}$ (see Figs. 2c,d of 
Boss 2001a), suggesting that water will be in a solid form, at
least initially. The question of whether water and other ices will 
remain solid during the core formation phase remains to be investigated
in detail. If the ices sublimate, icy core formation could be halted (the
critical temperature for water ice is 647 K at a pressure of 218 atm). 
Addressing this issue will require a detailed two-fluid calculation
of the heating and cooling processes in a protoplanet during the
core formation process. At any rate, more refractory species should
continue to settle to the core; Uranus's core is thought to be composed
of 2/3 ice and 1/3 rock (Hubbard 1997).

\page
\centerline{\bf PHOTOEVAPORATION}
\medskip

 An isolated giant gaseous protoplanet should contract toward planetary
densities on a time scale of $\sim 10^5$ yrs (Bodenheimer {\it et al.} 
1980, Cameron {\it et al.} 1982). We hypothesize that during this period, 
a massive star formed in the same star-forming region as the solar nebula, 
and began to produce copious EUV irradiation, similar to the situation 
in the Orion Nebula cluster today. OB associations like Orion appear 
to be the birthplace of most stars (Bally {\it et al.} 1998),
so an Orion-like environment should have influenced the formation 
of many planetary systems, possibly including our own. Significant losses 
(Richling and Yorke 1998) of disk gas can occur in as little as $\sim 10^4$ 
yrs for the protoplanetary disks (proplyds) close to the Orion Trapezium 
star $\Theta^1$ C Ori, while disk survival times are estimated (Bally 
{\it et al.} 1998) to be of order $10^5$ yrs, depending on the disk mass.
For a disk massive enough to undergo a gravitational instability, 
the survival time may be $\sim 10^6$ yrs (see below). Note that
we have assumed here that the disk instability which forms the
giant protoplanets occurs prior to the onset of EUV irradiation,
but this need not necessarily be the case. Because external EUV radiation
photoevaporates a disk from the outside in, it is conceivable that at the
optically thick disk midplane, the disk may remain sufficiently cool 
(25 K to 50 K) for a disk instability to proceed even after irradiation
has commenced.

 EUV irradiation, however, cannot evaporate disk gas closer
than about 5 to 10 AU from a solar-mass protostar, because the
gas temperatures produced ($\sim 10^4$ K) are insufficient to
permit a thermally-driven wind to exit the gravitational
potential well of the star (p. 761 of Johnstone {\it et al.} 1998).
This hot gas will remain gravitationally
bound to the star, forming a protective, roughly spherical halo with a radius 
of about 5 to 10 AU. Hence we expect that protoJupiter will remain 
shielded from the EUV flux by the persistence of the inner solar nebula
inside this halo, protoSaturn will be only partially shielded by 
the halo, while protoUranus and protoNeptune, orbiting outside the halo,
will soon become unshielded as the disk evaporates from the outside in,
and so they will begin to lose their gaseous envelopes. The proto-ice
giants evidently need to lose $\sim$ 2 Jupiter masses ($M_J$) of gas (Table 1),
with the amount needed to be removed decreasing with decreasing orbital
distance and falling to zero inside 10 AU. The mass loss rate for gas 
removal from disks or clumps close ($\sim 10^{17}$ cm) to $\Theta^1$ C Ori 
by EUV irradiation is estimated (Johnstone {\it et al.} 1998) to be

$$ \dot M \sim 8 \times 10^{-23} \ r_c \ M_\odot yr^{-1}, $$

\noindent
where $r_c$ is the radius of the clump in cm. The radii of giant
gaseous protoplanets are limited by tidal truncation by the
protosun, with the critical tidal radius being proportional
to orbital distance (Boss 1998), so disk instability models (Boss 2001a) 
showing the formation of clumps with radii of $\sim 0.6$ AU at 10 AU
would predict clumps with radii of $\sim$ 1.5 AU at a 25 AU distance.
Such a clump would then lose mass at a rate of $\sim 2 \times
10^{-6} M_J$ yr$^{-1}$, sufficient to lose the required amount
of envelope gas in $\sim 1$ Myr. This mass loss rate may also
be used to estimate the time scale for EUV radiation to first remove
the disk gas ($\sim 0.04 M_\odot$) residing between 20 AU and 30 AU 
in a marginally gravitationally unstable disk, which is $\sim 1$ Myr.

 While an isolated clump may contract to considerably smaller
radii than 1.5 AU in about $10^5$ yrs (Bodenheimer {\it et al.} 1980,
Cameron {\it et al.} 1982), once EUV radiation reaches the 
protoplanet, it will be embedded in a high temperature (10,000 K) plasma 
that may slow the escape of its radiation, stop any further 
contraction, and may even expand the protoplanet as its outer 
layers are being stripped away (Cameron {\it et al.} 1982).
In fact, once gas giant protoplanets form, they are likely to 
accrete or otherwise clear the disk gas in their vicinity, opening
up disk gaps, though this phase of evolution has not been encountered
in the simulations to date. Unless the OB star lies exactly in the disk plane
and EUV light cannot be scattered onto the protoplanets,
the star's EUV radiation should begin to strip the outer layers
of the protoplanets immediately, well before the unaccreted disk 
gas has been photoevaporated. Hence the estimated time scale of $\sim 1$ Myr 
for protoplanet envelope removal may still be approximately correct,
though it remains to be seen whether the onset of photoevaporation
in the outermost layers can significantly slow the contraction
of a protoplanet. If not, then the OB star might have to be located
closer to the disk than is assumed here, nominally $10^{17}$ cm, in order
to have envelope removal occur on a time scale comparable to that for
protoplanet contraction. 

 On the other hand, it must be noted that the protoplanet contraction time 
estimates of Bodenheimer {\it et al.} (1980) and Cameron {\it et al.} 
(1982) are based on spherically symmetric models, where rotation is ignored.
In reality, gas giant protoplanets will be rotating very rapidly as they
form (Boss 2001a), and their rate of contraction will be limited not only 
by the rate at which they can radiate the compressional energy produced
by contraction, but also by the rate at which they can transfer angular 
momentum out of their central regions. These protoplanets should thus look
like rotationally-flattened polytropes, with outer regions where centrifugal 
forces provide significant support against self-gravity. Because when the 
protoplanets first form, they are only marginally self-gravitating
(Boss 2001a), and because of the heating associated with further contraction 
toward planetary densities, these protoplanets should have gravitational 
instability parameters $Q > 1.5$ during their evolution, and thus should not
experience sub-fragmentation. It is unclear what processes will control
the outward transport of angular momentum in these objects, but it is certain 
that the time scale for contraction must be longer than that in the absence 
of rotation. Hence contraction times based on spherically symmetric
models should be lower limits, and the true contraction times should
permit significantly more time for EUV to photoevaporate the rotationally
distended envelopes of the outer protoplanets. Detailed modeling of the
evolution of a rotating gaseous protoplanet is needed to address 
this key issue.

 EUV radiation will not be able to remove the entire gaseous envelope
of protoUranus and protoNeptune, for the same reason that the solar
nebula inside 5 to 10 AU cannot be removed. The inner radius $r_g$ 
of EUV removal (Johnstone {\it et al.} 1998) is 

$$ r_g \approx {1 \over 2} {G M_p \over c_s^2},$$

\noindent
where $M_p$ is the planet mass and $c_s = 10$ km s$^{-1}$ for gas heated
by EUV radiation. For a final planet mass of $\sim 15 M_\oplus$, 
$r_g = 3 \times 10^9$ cm, slightly larger than the present-day
radii of Uranus and Neptune of $\sim 2.5 \times 10^9$ cm, implying
that EUV envelope stripping might stop short of the solid core,
leaving behind a small gaseous envelope, as is observed. 
Other choices of $M_p$ would yield other values of $r_g$; this
analysis simply shows that $M_p \sim 15 M_\oplus$ is consistent with EUV
irradiation and the present radii of the ice giant planets.
Saturn appears to be an intermediate case between the envelope losses
suffered by the ice giants and the envelope retention enjoyed by
Jupiter, due to its location close the point where the solar
nebula could no longer be evaporated by external EUV radiation.

 Removal of most of the gaseous envelopes of protoUranus and protoNeptune
thus requires $\sim 1$ Myr of sustained EUV irradiation from a
nearby massive star. Is this likely? The lifetimes of massive 
($> 8 M_\odot$) stars can exceed $\sim$ 10 Myr, and the
ionizing stars of the Trapezium cluster (Prosser {\it et al.} 1994)
have already lived for 0.3 to 2 Myr. Star-forming OB 
associations (Elmegreen {\it et al.} 2000) have lifetimes on the 
order of 10 Myr, so $\sim 1$ Myr of sustained EUV irradiation from an 
external source seems quite possible. The solar nebula need not have 
remained at a fixed distance from a star similar to $\Theta^1$ C Ori during 
EUV irradiation: Scally and Clarke (2001) have shown that stars orbiting 
throughout the Orion Nebula cluster will lose considerable disk mass
(0.01 to 1 $M_\odot$) by UV photoevaporation within about 2 Myr.
The early Sun itself (Shu {\it et al.} 1993) may have produced enough
EUV radiation to remove the outer solar nebula gas beyond $\sim$ 9 AU
over a time of $\sim 10^7$ yrs, but protoUranus and protoNeptune
would have contracted to planetary-size during this period, making
EUV evaporation inefficient. A strong external source of EUV radiation 
seems to be required for our scenario to work. 

\page
\centerline{\bf CONCLUSIONS}
\medskip

 While we have focused on EUV ionizing radiation ($\lambda < 912 \dot A$),
far-ultraviolet (FUV) radiation ($\lambda > 912 \dot A$) can also play
an important role in evaporating protoplanetary disks (Johnstone {\it et al.} 
1998). The four luminous Trapezium stars provide the EUV and FUV photons
which bathe the protostars in the Orion Nebula cluster. FUV radiation
will aid in the removal of gaseous envelopes from the proto-ice giants.

 Because most stars form in OB associations, EUV and FUV irradiation may play a 
much more important role in planetary system formation than has previously
been thought to be the case. As we have shown here, EUV and FUV 
irradiation need not prevent the formation of gas or ice giant planets.
Given that the emerging results (Butler {\it et al.} 2000, Boss 2001b)
from the ongoing census 
of extrasolar gas giant planets give no indications that gas giant 
planets are rare, it seems that star-forming regions like Orion must 
be capable of forming at least gas giant planets, and presumably
entire Solar Systems.

\medskip
\centerline{\bf APPENDIX: ORBITAL STABILITY PRIOR TO IRRADIATION}
\medskip

 Considering that the giant protoplanets required in Table 1 are more
massive than the existing giant planets, one question which arises
regards the stability of their orbits -- can these protoplanets
remain on stable orbits long enough for EUV irradiation to reduce their
masses to their current values? The orbital stability analysis of
Chambers {\it et al.} (1996) suggests that systems of more
than two planets (with masses $m_1, m_2$) will be orbitally unstable 
around a 1 $M_\odot$ star if their semimajor axes ($a_1, a_2$) are 
separated by less than about 10 $R_H$, where the mutual Hill radius is 
defined to be $R_H = [(m_1 + m_2)/(3 M_\odot)]^{1/3} (a_1 + a_2)/2$.
For the four giant planets shown in Table 1, located at the current
orbital distances of the outer planets, all three pairs of planets 
are located closer together than 10 $R_H$ (5.9 $R_H$, 6.1 $R_H$, and 
3.9 $R_H$, respectively, moving outward), implying instability. 
However, the time scale for the system to disrupt is estimated to be
of the order of 10 Myr for the innermost three giant planets, with
separations of $\sim 6 R_H$ (extrapolating the results shown in 
Figure 3 of Chambers {\it et al.} 1996 to three planets, each with a mass
$\sim 10^{-3} M_\odot$), though the outermost planet has a smaller
separation (3.9 $R_H$) from its adjacent planet, which should shorten this
time scale. In order to further test the stability, one of us (NH) has 
used the SWIFT symplectic integrator code of Levison and Duncan (1994)
to integrate the three dimensional orbits of four planets in orbit around
the Sun, starting with the current positions of the four outer planets 
(CHO: Cohen {\it et al.} 1973) and with the masses given in 
Table 1. With the CHO initial conditions, the system is stable for at 
least 10 Myr; only relatively small changes ($\sim$ 10\%) in semimajor axes 
are evident during this period. For other arbitrary choices of the initial 
conditions, however, the time scale for orbital instability can be as 
short as a few 0.1 Myr. The SWIFT calculations do not include the
interactions of the planets with the disk gas, and so cannot be
considered definitive -- for example, the protoplanets may have formed
with larger mutual separations and then undergone orbital migration due to
disk interactions, prior to or during photoevaporation of the outer disk 
gas. While the orbital stability of the nominal initial configuration 
implied by Table 1 is an important question to consider in future work,
we believe that the SWIFT integration from the CHO initial conditions
suggests that orbital stability for several Myr or more is at least
plausible, which is a time period sufficiently long for EUV irradiation to 
lower the ice giant protoplanet masses to their present values.

\medskip
\centerline{\bf ACKNOWLEDGMENTS}
\medskip

 We thank Richard Durisen and Tristan Guillot for valuable referee 
reports. APB thanks Kathy Sawyer for questions about planet formation in
the Orion Nebula. This work was partially supported by the NASA 
Planetary Geology and Geophysics Program under grant NAG5-10201
and by the NASA Origins of Solar Systems Program under grant NAG5-10547.

\page
\centerline{\bf REFERENCES}
\parindent=-5pt
\medskip

Bally, J., L. Testi, A. Sargent, and J. Carlstrom 1998. Disk mass limits
and lifetimes of externally irradiated young stellar objects embedded
in the Orion Nebula. {\it Astron. J.} {\bf 116}, 854-859.

\smallskip

Bodenheimer, P., A. S. Grossman, W. M. DeCampli, G. Marcy, 
and J. B. Pollack 1980. Calculations of the evolution of the giant planets.
{\it Icarus} {\bf 41}, 293-308.

\smallskip

Boss, A. P. 1998. Evolution of the solar nebula. IV. Giant gaseous protoplanet 
formation. {\it Astrophys. J.} {\bf 503}, 923-937.

\smallskip

Boss, A. P. 2000. Possible rapid gas giant planet formation in the solar 
nebula and other protoplanetary disks. {\it Astrophys. J.} {\bf 536},
L101-L104.

\smallskip

Boss, A. P. 2001a. Gas giant protoplanet formation: disk instability
models with detailed thermodynamics and radiative transfer.
{\it Astrophys. J.}, in press.

\smallskip

Boss, A. P. 2001b. Giant giants or dwarf dwarfs? {\it Nature} {\bf 409},
462-463.

\smallskip

Brice\~no, C. and 10 colleagues 2001. The CIDA-QUEST large-scale survey of 
Orion OB1: Evidence for rapid disk dissipation in a dispersed stellar 
population. {\it Science} {\bf 291}, 93-96.

\smallskip

Bryden, G., D. N. C. Lin, and S. Ida 2000. Protoplanetary formation.
I. Neptune. {\it Astrophys. J.} {\bf 544}, 481-495.

\smallskip

Butler, R. P., S. S. Vogt, G. W. Marcy, D. A. Fischer, G. W. Henry,
and K. Apps 2000. Planetary companions to the metal-rich stars
BD -10$^o$3166 and HD 52265. {\it Astrophys. J.} {\bf 545}, 504-511.

\smallskip

Cameron, A. G. W., W. M. DeCampli, and P. Bodenheimer 1982. Evolution of 
giant gaseous protoplanets embedded in the primitive solar nebula.
{\it Icarus} {\bf 49}, 298-312.

\smallskip
 
Canup, R. M., and K. Righter (eds.) 2000. {\it Origin of the Earth and 
Moon}, 555 pages, Univ. of Arizona Press, Tucson.

\smallskip

Chambers, J. E., G. W. Wetherill, and A. P. Boss 1996. The Stability
of Multi-Planet Systems. {\it Icarus} {\bf 119}, 261-268.

\smallskip

Cohen, C. J., E. C. Hubbard, and C. Oesterwinter 1973. Elements
of the outer planets for one million years. {\it Astron. Pap.
Am. Ephem.} {\bf 22}, 3 (CHO).

\smallskip
 
Elmegreen, B. G., Y. Efremov, R. E. Pudritz, and H. Zinnecker 2000.
Observations and theory of star cluster formation. In {\it Protostars
and Planets IV} (V. Mannings, A. P. Boss, and S. S. Russell, Eds.), pp.
179-215. Univ. of Arizona Press, Tucson.

\smallskip

Guillot, T. 1999. A comparison of the interiors of Jupiter and Saturn.
{\it Planet. Space Sci.} {\bf 47}, 1183-1200.

\smallskip

Hubbard, W. B. 1992. Planetary interiors, Jovian planets. In 
{\it The Astronomy and Astrophysics Encyclopedia} (S. P. Maran, Ed.), pp.
525-527. Van Nostrand Reinhold, New York.

\smallskip

Hubbard, W. B. 1997. Uranus. In {\it Encyclopedia of Planetary Sciences} 
(J. H. Shirley and R. W. Fairbridge, Eds.), pp. 856-857. Chapman
and Hall, London.

\smallskip

Johnstone, D., D. Hollenbach, and J. Bally 1998. Photoevaporation of
disks and clumps by nearby massive stars: application to disk
destruction in the Orion Nebula. {\it Astrophys. J.} {\bf 499}, 758-776.

\smallskip

Kokubo, E., and S. Ida 1998. Oligarchic growth of protoplanets. {\it Icarus}
{\bf 131}, 171-178.

\smallskip

Kortenkamp, S. J., and G. W. Wetherill 2000. Terrestrial planet and 
asteroid formation in the presence of giant planets. I. Relative
velocities of planetesimals subject to Jupiter and Saturn perturbations.
{\it Icarus} {\bf 143}, 60-73.

\smallskip

Kortenkamp, S. J., G. W. Wetherill, and S. Inaba 2001. Runaway growth
of planetary embryos facilitated by massive bodies in a protoplanetary
disk. {\it Science} {\bf 293}, 1127-1129.

\smallskip

Laughlin, G., and P. Bodenheimer 1994. Nonaxisymmetric evolution in
protostellar disks. {\it Astrophys. J.} {\bf 436}, 335-354.

\smallskip

Levison, H. F., and M. J. Duncan 1994. The Long-Term Dynamical Behavior
of Short-Period Comets. {\it Icarus} {\bf 108}, 18-36.

\smallskip

Levison, H. F., and G. R. Stewart 2001. Note: Remarks on modeling the formation
of Uranus and Neptune. {\it Icarus}, in press.

\smallskip

Lissuaer, J. J., J. B. Pollack, G. W. Wetherill, and D. J. Stevenson 1995.
Formation of the Neptune system. In {\it Neptune} (D. P. Cruikshank, Ed.),
pp. 42-59. Univ. of Arizona Press, Tucson.

\smallskip

Nelson, A. F., W. Benz, F. C. Adams, and D. Arnett 1998. Dynamics of
circumstellar disks. {\it Astrophys. J.} {\bf 502}, 342-371.

\smallskip

Nelson, A. F., W. Benz, and T. V. Ruzmaikina 2000. Dynamics of
circumstellar disks. II. Heating and cooling. {\it Astrophys. J.} 
{\bf 529}, 357-390.

\smallskip

Pickett, B. K., P. Cassen, R. H. Durisen, and R. Link 2000a.
The effects of thermal energetics on three-dimensional hydrodynamic
instabilities in massive protostellar disks. II. High-resolution
and adiabatic evolutions. {\it Astrophys. J.} {\bf 529}, 1034-1053.

\smallskip

Pickett, B. K., R. H. Durisen, P. Cassen, and A. C. Mejia 2000b.
Protostellar disk instabilities and the formation of substellar
companions. {\it Astrophys. J.} {\bf 540}, L95-L98.

\smallskip

Pollack, J. B., D. Hollenbach, S. Beckwith, D. P. Simonelli, T. Roush,
and W. Fong 1994. Composition and radiative properties of grains in
molecular clouds and accretion disks. {\it Astrophys. J.} {\bf 421}, 
615-639.

\smallskip

Pollack, J. B., O. Hubickyj, P. Bodenheimer, J. J. Lissauer, M. Podolak,
and Y. Greenzweig 1996. Formation of the giant planets by concurrent
accretion of solids and gas. {\it Icarus} {\bf 124}, 62-85.

\smallskip

Prosser, C. F., J. R. Stauffer, L. Hartmann, L., D. R. Soderblom, 
B. F. Jones, M. W. Werner, and M. J. McCaughrean 1994. {\it HST} photometry
of the Trapezium cluster. {\it Astrophys. J.} {\bf 421}, 517-541.

\smallskip

Richling, S., and H. W. Yorke 1998. Photoevaporation of protostellar disks.
IV. Externally illuminated disks. {\it Astron. Astrophys.} {\bf 340},
508-520.

\smallskip

Scally, A., and C. Clarke 2001. Destruction of protoplanetary discs
in the Orion Nebula Cluster. {\it Mon. Not. Roy. Astr. Soc.} {\bf 325},
449-456.

\smallskip

Shu, F. H., D. Johnstone, and D. Hollenbach 1993. Photoevaporation of the
solar nebula and the formation of the giant planets. {\it Icarus}
{\bf 106}, 92-101.

\smallskip

Thommes, E. W., M. J. Duncan, and H. F. Levison 1999. The formation of
Uranus and Neptune in the Jupiter-Saturn region of the Solar System.
{\it Nature} {\bf 402}, 635-638.

\smallskip

Weidenschilling, S. J. 1988. Formation processes and time scales for
meteorite parent bodies. In {\it Meteorites and the Early Solar
System} (J. F. Kerridge and M. S. Matthews, Eds.), pp. 348-371. Univ. 
of Arizona Press, Tucson.

\smallskip

Weidenschilling, S. J., and J. N. Cuzzi 1993. Formation of planetesimals
in the solar nebula. In {\it Protostars and Planets III},
(E. H. Levy and J. I. Lunine, Eds.), pp. 1031-1060. Univ. 
of Arizona Press, Tucson.

\smallskip

Wuchterl, G. 1993. The Critical Mass for Protoplanets Revisited: Massive
Envelopes through Convection. {\it Icarus} {\bf 106}, 323-334.

\smallskip

Wuchterl, G., T. Guillot, and J. J. Lissauer 2000. Giant planet formation. 
In {\it Protostars and Planets IV} (V. Mannings, A. P. Boss, and S. S. 
Russell, Eds.), pp. 1081-1109. Univ. of Arizona Press, Tucson.

\smallskip

\vfill\eject
\centerline{\bf TABLE 1}
\parindent=0pt
\medskip

Total, inferred core (Guillot 1999, Hubbard 1992)
and resulting envelope masses (in Earth masses, 
$M_\oplus$) for the gas and ice giant planets, masses of the gas giant 
protoplanets (GGPP, in Jupiter masses, $M_J$) needed to form the observed 
cores by the disk instability mechanism, and masses of envelope gas that must
be removed by EUV irradiation (in $M_J$) to yield the observed planet.

\settabs\+ Mass Mass Mass \quad \quad
& \quad \quad Jupiter \quad \quad 
& \quad \quad Saturn \quad \quad 
& \quad \quad Uranus \quad \quad
& \quad \quad Neptune \quad \quad 
\cr

\medskip
\hrule
\medskip

\+ Mass & Jupiter & Saturn & Uranus & Neptune \cr

\medskip
\hrule
\medskip

\+ Total ($M_\oplus$) &  318 &  95 &  15 &  17 \cr

\smallskip

\+ Core ($M_\oplus$) & 6 & 12 & 12 & 16 \cr

\smallskip

\+ Envelope ($M_\oplus$) & 312 & 83 & 3 & 1 \cr

\smallskip

\+ GGPP ($M_J$) & 1 & 2 & 2 & 2.5 \cr

\smallskip

\+ Removed ($M_J$) & 0 & 1.7 & 1.95 & 2.45 \cr

\medskip
\hrule
\medskip

\bye